# Production and mechanical characterization of graphene micro-ribbons

Maria Giovanna Pastore Carbone[1], Georgia Tsoukleri[1], Anastasios C. Manikas[1,2], Eleni Makarona[3], Christos Tsamis[3] and Costas Galiotis[1,2,*]

1. Institute of Chemical Engineering Sciences, Foundation of Research and Technology- Hellas (FORTH/ICE-HT), Stadiou Street, Platani, Patras, 26504 Greece.
2. Department of Chemical Engineering, University of Patras, Patras, 26504 Greece
3. Institute of Nanoscience and Nanotechnology, National Center for Scientific Research "Demokritos", Athens 15310, Greece
* Correspondence: c.galiotis@iceht.forth.gr; Tel.: +302610-965255

Abstract: Patterning of graphene into micro- and nano-ribbons allows for the tunability in emerging fields such as flexible electronic and optoelectronic devices, and is gaining interest for the production of more efficient reinforcement for composite materials. In this work we fabricate micro-ribbons from CVD graphene by combining UV photolithography and dry etching oxygen plasma treatments. Raman spectral imaging confirms the effectiveness of the patterning procedure, which is suitable for large-area patterning of graphene on wafer-scale, and confirms that the quality of graphene remains unaltered. The produced micro-ribbons were finally transferred and embedded into a polymeric matrix and the mechanical response was investigated by in-situ mechanical investigation combining Raman spectroscopy and tensile/compressive tests.

Keywords: graphene; micro-ribbon; UV lithography; mechanical properties; Raman spectroscopy

1. Introduction

Graphene is a perfect 2D crystal of covalently bonded carbon atoms that is a promising candidate in a number of electrical, thermal and mechanical applications, due to its exceptional physical properties[1]. In particular, due to outstanding transport properties, graphene is the best candidate for the production of transparent electrodes and optoelectronic devices[2]. The production of graphene sheets on large scale can be obtained from both physical and chemical routes, such as epitaxial growth on silicon carbide, chemical exfoliation and reduction of graphene oxide, as well as the chemical vapor deposition (CVD)[3, 4]. The mechanical cleavage of highly oriented pyrolytic graphite (HOPG) produces pristine graphene with low concentration of defects but the size, the thickness and the location of the flakes cannot be controlled. Graphene derived from chemical processing are usually irregular in shape and are not suitable for integrated circuit or microdevice manufacturing. The CVD method yields reliable production of large-area and high-quality graphene films on a variety of metal substrates at relatively low cost. For practical applications in device manufacturing, however, several patterning techniques, such as optical lithography or electron beam lithography followed by lift off and plasma etching procedures, or femtosecond laser etching process, are required to tailor the graphene sheets into desired sizes and shapes [5, 6]. One example is the production of graphene micro-ribbons and nano-ribbons as waveguiding strain sensors[7], chemiresistors[8], microsupercapacitors[9], and tunable tetrahertz metamaterials [10].

Graphene also attracts great attention as strengthening component in composites, as it is attributed a unique combination of mechanical properties in tension[11]. In this regard, the regular geometry of the micro-ribbons is expected to ensure a uniform stress distribution[12] as well as to prevent the lateral buckling effect which is commonly induced by the Poisson's contraction in thin membranes and has been observed in supported graphenes[13]. In light of the above, graphene micro-ribbons could be exploited also as more efficient reinforcement in composite materials. However, to the best of our knowledge, experimental studies focused on the investigation of the mechanical performance of graphene micro-ribbon embedded in polymeric have not been reported to date in the literature.

The performance of graphene-based devices and of graphene-reinforced composites is strongly influenced by the interfacial properties between graphene and substrate materials. Hence, for applications that involve stretching or bending of the substrate, such as in flexible electronic devices, and in structural composites, it is essential to know whether the stress is efficiently transferred from the matrix/ substrate to the graphene flakes. Raman spectroscopy has been successfully adopted to investigate deformation mechanisms of graphene and stress transfer in a variety of graphene-reinforced composites[14]. Actually, carbon-based materials show well-defined Raman spectra and their Raman bands are found to shift under the imposition of stress thus enabling stress-transfer between the matrix and reinforcing phase to be monitored[15].

In this work, the fabrication of micro-ribbons from large area graphene film synthesized by CVD method with UV lithography and oxygen plasma process is discussed. The design of the micro-ribbons has been accurately performed taking into account some important experimental findings by our group[16]. Raman spectral imaging and AFM have been adopted to evaluate the effectiveness of the micropatterning process on wafer-scale and the quality of the final micro-ribbons. Finally, the produced micro-ribbons were transferred and embedded into a polymeric matrix and their mechanical response in tension and compression was investigated by simultaneous Raman spectroscopy.

2. Materials and Methods

Graphene produced via CVD on 3-inch copper/silicon (Cu/Si) wafers was gently supplied by AIXTRON SE (Herzogenrath, Germany). Graphene micro-ribbons were produced directly on the Si/Cu wafer via UV standard lithography and oxygen plasma etching methods. The overall process for the production of graphene micro-ribbons and the transfer onto a polymer bar is schematized in Fig. 1a. After the growth of graphene monolayer on a Cu/Si wafer, a thin layer (~1μm thickness) of a photoresist AZ 726MIF was spin coated on top at 5000 rpm for 30 sec and then was baked for 10min at 95oC. Afterwards, the mask was loaded to the mask aligner and a positive UV standard lithographic process was performed for 7 sec, then the sample was developed in AZ 726MIF developer solution for 1min, rinsed with DI water and dried gently with nitrogen. Next, the wafer was placed in the etching chamber for 30 sec and, by following the below conditions: RF frequency = 13.56 MHz, RF power = 50 Watt, maximum vacuum = 10-4 mbar, pressure = 100 mTorr and oxygen N55 flow = 50 sccm, the uncovered graphene area was dry etched. Finally, acetone and 2-propanol were used to remove the remaining photoresist from the covered ribbons and the sample was rinsed with DI water and was dried gently with nitrogen.

The produced graphene micro-ribbons were finally embedded into a poly (methyl methacrylate) (PMMA) substrate by using the "wet transfer" method (Fig. 1b). According to this method, a thin transfer PMMA layer (~500nm thickness) of 495 K molecular weight dissolved in anisole (MicroChem) was first spin coated on top of the surface of the graphene micro-ribbons/Cu/Si wafer; then, the copper layer was etched in 1M solution of ammonium persulphate (APS). PMMA/graphene micro-ribbons membrane was floating on the solution and then was rinsed in DI water. Finally, the PMMA/graphene micro-ribbons thin film was "fished" by using a PMMA beam. During transfer process, care has been taken to align the micro-ribbons to the PMMA beam axis.

Raman spectroscopy was used to evaluate the quality of the produced micro-ribbons. Raman spectra were recorded with the InVia Renishaw Raman system, under a 100x objective, by using an Argon ion laser (514.5 nm) and a diode laser (785 nm) as the excitation sources. Laser power on sample was kept below 0.25 mW in order to avoid laser-induced heating. A motorized xy stage allowed the collection of a large amount Raman spectra across the sample with a step size of 1 μm. By using the cantilever beam technique, uniaxial tensile and compressive loadings were applied to graphene micro-ribbons embedded into the PMMA, collecting simultaneously Raman spectra in the middle of the ribbons. The beam was bended first in tension by steps of 0.1%, and then in compression by steps of 0.05%, by simply changing the bending direction. The polarization of the incident light was kept parallel to the applied strain axis. Detailed description of the apparatus and of sample geometry are reported elsewhere[16, 17]. Atomic force microscopy (AFM) was carried out with a commercial AFM (Dimension icon, Bruker Co., USA) on micro-ribbons supported on PMMA films (i.e., before fishing procedure). The ScanAsyst mode was applied using silicon nitride tip (Scanasyst-Air, Bruker, nom. tip radius 2 nm, nom. frep. 70 kHz, nom. spring constant of 0.4 N/m) with scan resolution of 512 samples per line. Height mode images were acquired simultaneously at a fixed scan rate (0.1 Hz) with a resolution of 512 × 512 pixels.

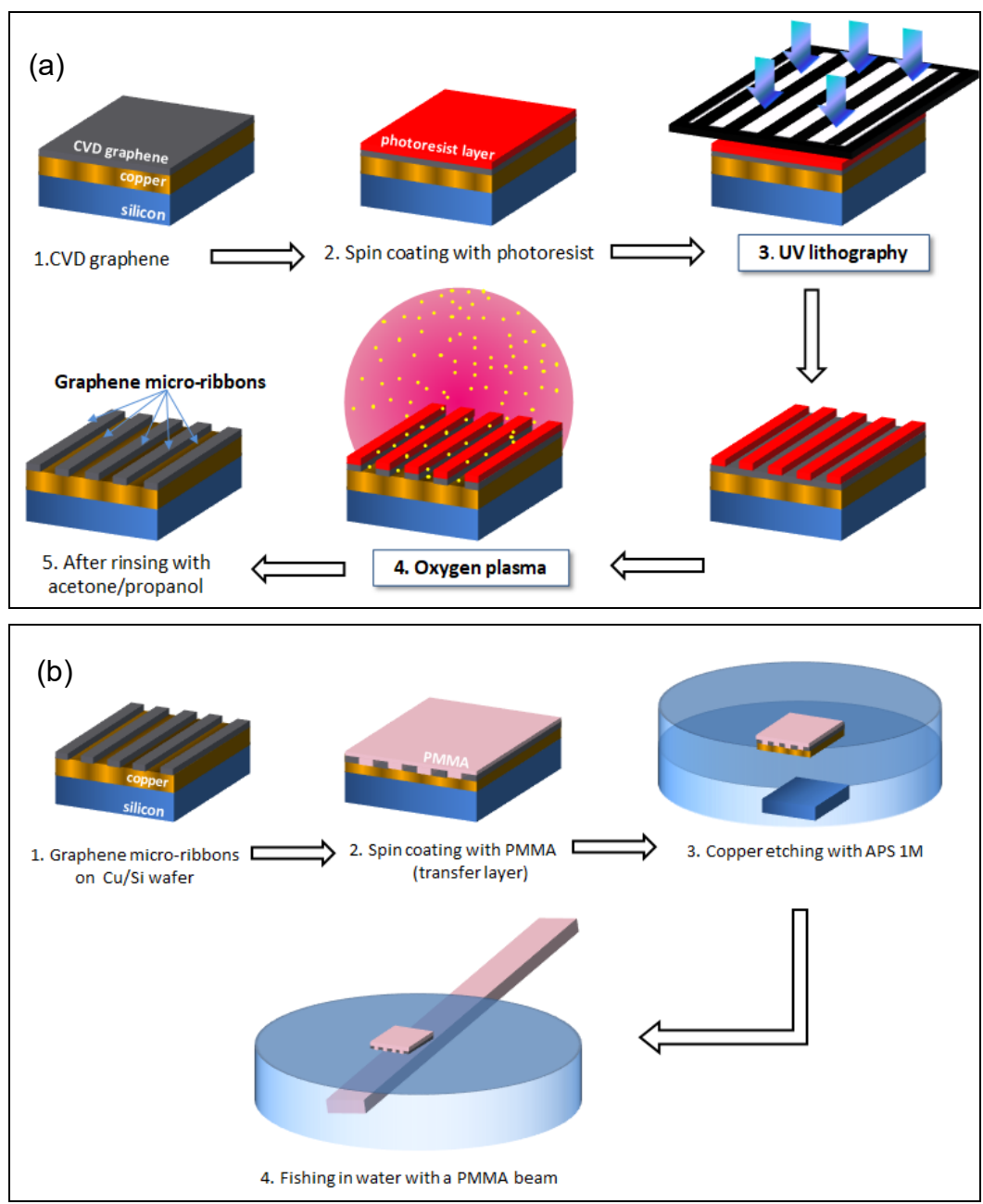

Figure 1. Micro-patterning of CVD graphene micro-ribbons (a) and transfer onto polymeric matrix (b)

## 3. Results and Discussion

### 3.1. CVD graphene-microribbons

After photolithography/plasma process on the Gr/Cu/Si wafer, the fabricated graphene micro-ribbons are clearly observed under an optical microscope, due to the presence of the photoresist on top of them (Fig. 2a) and detailed Raman mappings were conducted using 514.5 nm excitation wavelength. A characteristic contour mapping is presented in Fig. 2b, where the blue areas correspond to produced graphene micro-ribbons of several widths from 1 to 4.5 μm. As can been seen, the contour is similar to the pattern of the designed basic unit, verifying that the above described production process was successfully performed.

After transferring on a PMMA bar, the effectiveness of the whole fabrication process and the quality of the embedded graphene micro-ribbon/PMMA samples have been investigated by using detailed Raman maps. The characteristic contour maps of frequency positions and of full width at half maximum (FWHM) of both G and 2D peaks are given in Fig. 3. The micro-ribbon appears well aligned to PMMA beam axis and shows a large distribution in the peak position and the FWHM of 2D and G peaks, which points to a not perfectly flat micro-ribbon thus suggesting the presence of structural faults such as wrinkles, folds, substrate surface defects and molecular trapping induced by the fabrication and transferring process.

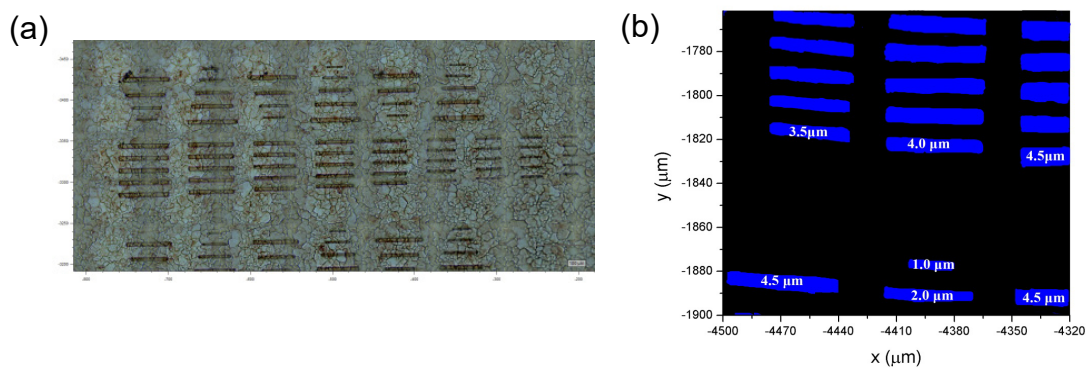

Figure 2. a) Optical image of produced graphene micro ribbons on Copper – Silicon wafer after Lithography and Plasma etching process. The micro – ribbons are still covered by the AZ 726MIF photoresist, b) Raman mapping on Graphene – Copper- Silicon wafer after dissolving AZ 726MIF photoresist. The blue colour corresponds to the intensity of 2D peak.

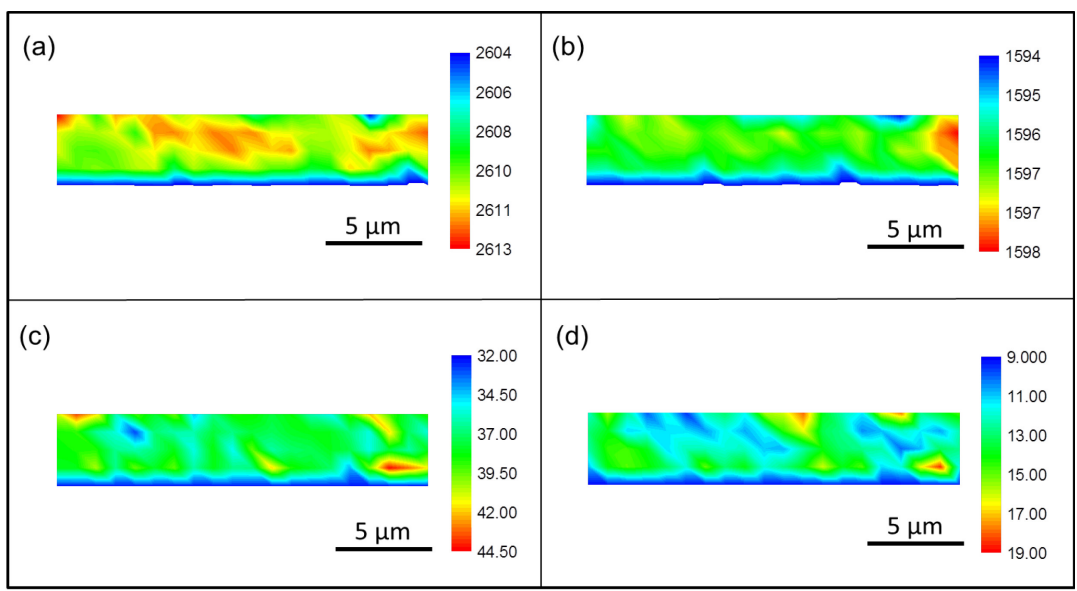

Figure 3. Characteristic contour maps of a) Pos(2D), b) Pos(G), c) FWHM(2D) and d) FWHM(G) of PMMA-embedded graphene micro-ribbon.

Table 1. Mean values and standard deviation of the position and FWHM of G and 2D bands for the PMMA-embedded micro-ribbons.

|  | Position (cm$^{-1}$) | FWHM (cm$^{-1}$) |
|---|---|---|
| G peak | 1596.2 ± 0.7 | 13 ± 2 |
| 2D peak | 2612 ± 1 | 38 ± 2 |

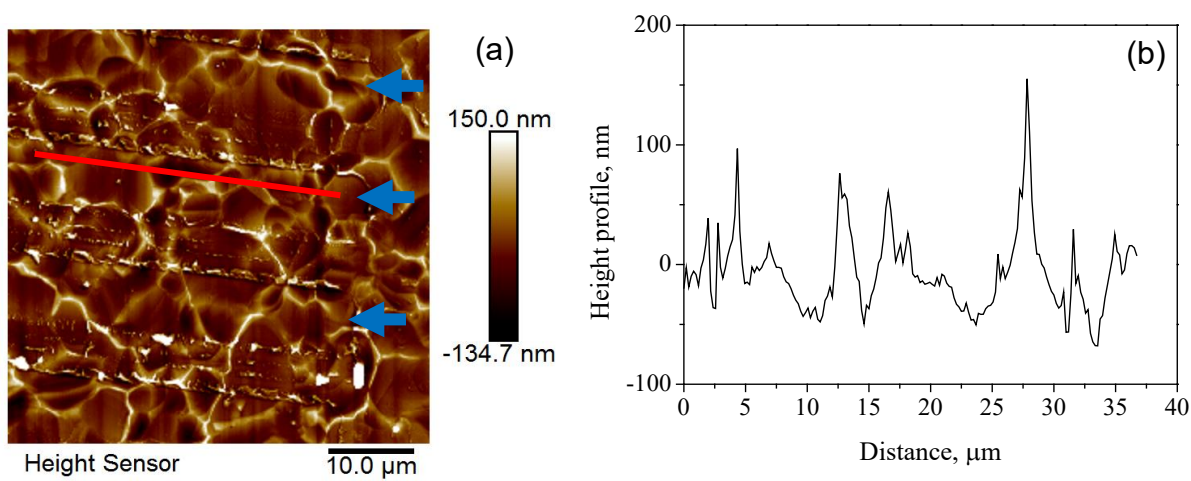

Figure 4. a) AFM image of a graphene micro-ribbons supported on PMMA and b) the corresponding height profile of the inset red line. Blue arrows indicate the micro-ribbons.

Statistical analysis of the maps has been performed and the mean values of positions and the FWHM of both 2D and G peaks are reported in Table 1. Based on the shift rates of the 2D and G Raman peaks for graphene flake under biaxial tensile deformations[18], the compressive strain experienced by the micro-ribbon can be estimated to be ~0.1 %. Furthermore, the average value of the ratio of the intensities of 2D and G peaks is ~1.4, likely suggesting chemical doping induced by the transfer process on the PMMA (see Figures S1 and S2). Also, the corrugation of graphene micro-ribbons supported on PMMA has been verified by AFM (in Figure 4), revealing that the structure of graphene micro-ribbons is highly crumpled and that the polymeric substrate is not flat.

Finally, the embedded micro-ribbons were subjected to tension and compression, and Raman spectra were collected at each strain level. Representative Raman spectra in the G and 2D peaks spectral regions of graphene at several strain levels are shown in Figure 5, and the fitted positions of the G and of the 2D band as a function of strain are reported in Figure 6. In tension, it has been found that the positions of both 2D and G peaks show a linear behaviour with strain; the strain sensitivity (from the linear part of the graph, red lines) of 2D is -14.8 cm$^{-1}$/%, while the strain sensitivity of G is -5.7 cm$^{-1}$/%. In compression, it has been found that the positions of both the 2D and G peaks exhibit a non-linear trend with strain. In particular, the average strain sensitivity in compression (from the linear part of the graph, red lines) of 2D is +12.3 cm$^{-1}$/%, while the strain sensitivity of G is +5.1 cm$^{-1}$/%. It should be noted here that G band splitting cannot be observed due to low deformation level experienced by the ribbons. These values of strain sensitivities are in agreement with those found for CVD graphene membranes deposited on polymeric substrates[19] and this confirms that the mechanical response of the micro-ribbons is not negatively affected after micro-patterning procedure. As observed by Li et al., the strain sensitivity of the 2D peak of a wrinkled CVD graphene was found to be less than 25% of the strain sensitivity of a flat exfoliated graphene flake[19]. This unusual Raman band shift behaviour was ascribed to the characteristic microstructure of CVD graphene - which is maintained in the graphene micro-ribbons (AFM image shown in Fig. 4a) - and was modelled in terms of mechanically isolated graphene islands separated by the graphene wrinkles.

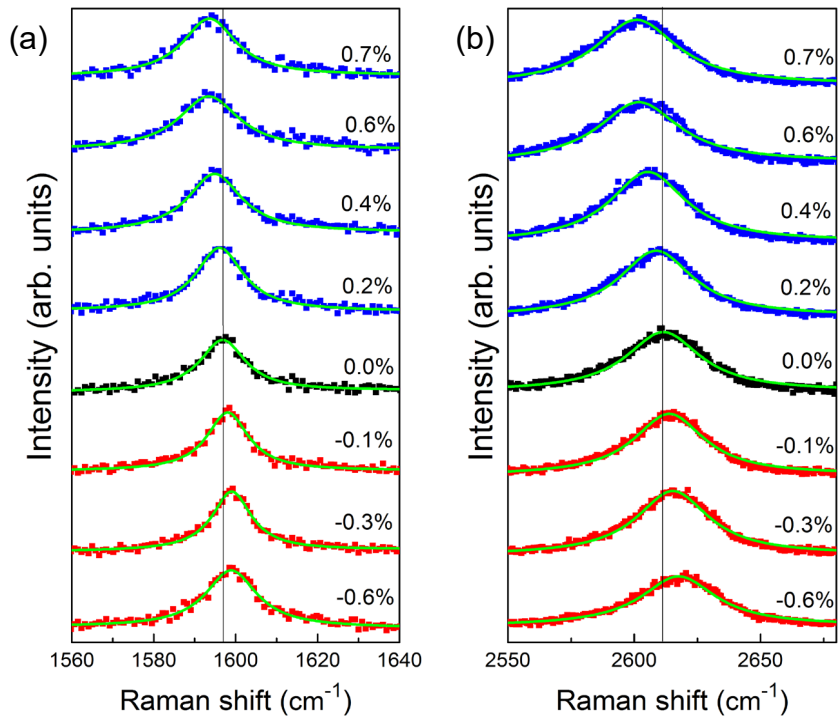

Figure 5. Evolution of the G (a) and 2D band (b) as a function of the strain, for the embedded CVD graphene micro–ribbons. The original measurements are plotted as points (blue for tension and red for compression). Solid lines represent Lorentzian fits to the experimental data.

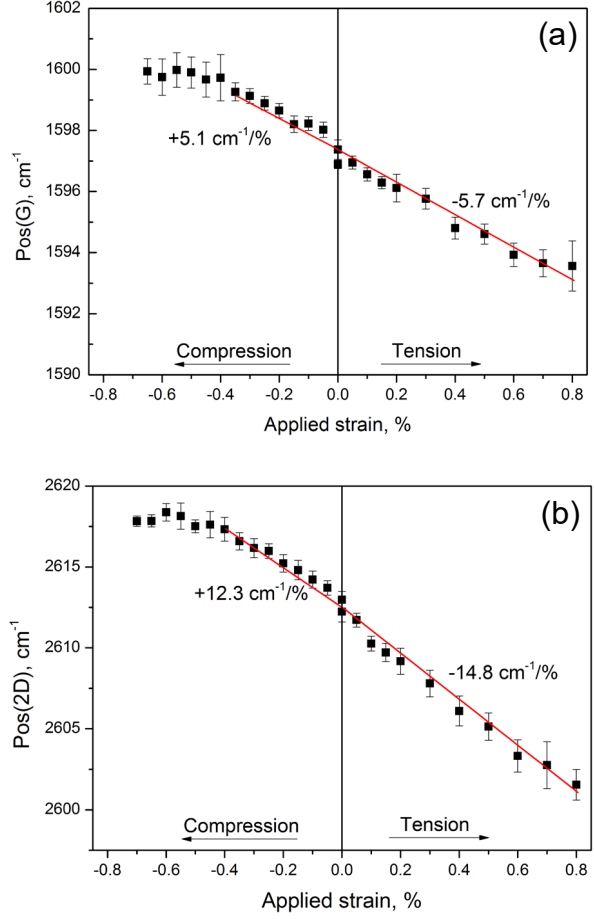

Figure 6. Position of G (a) and 2D (b) peak versus strain for the embedded CVD graphene micro–ribbons. Solid lines represent fits to the experimental data.

Furthermore, it is well established that the rate of Raman peak shift per unit strain of carbon-based materials can be related to their effective Young's modulus[15]. Therefore, since it has been demonstrated that graphene with a modulus of 1 TPa presents a 2D peak shift rate of -55 cm$^{-1}$/% strain[20] (for the 785 nm excitation line), it is then possible to evaluate the effective modulus of the graphene micro-ribbons embedded in the PMMA matrix from the measured 2D band shift data ($\frac{dPos(2D)}{d\varepsilon}$) using the following expression:

$$E_{eff} = -\frac{dPos(2D)}{d\varepsilon} \times \frac{1TPa}{-55} \quad (1)$$

By using the measured strain sensitivity in tension of the 2D peak, a value of effective Young's modulus of 270 GPa is estimated, which is in agreement with reported values of Young's modulus for wrinkled CVD graphene sheets[21]. Furthermore, by using the approach developed by Androulidakis et al. [20] we can convert the Raman wavenumber vs. strain curves to true axial stress by using the wavenumber per stress rates of -5.5 cm$^{-1}$ GPa$^{-1}$ (as determined for the 785 nm excitation). As seen, the data in tension are quite linear up to a value of strain of 0.8%, which was the limit of the experimental apparatus employed in our work. No material failure is observed up to that strain level that corresponds to a stress of ~2 GPa. In compression, monolayer graphene is expected to fail in air at very small strains (1 nanostrain) due to its almost zero thickness[16]. However, when embedded in polymer matrices both sides of the monoatomic membrane are prevented from out-of-plane deformation by the presence of the polymer. When the lateral van den Waals bonds eventually yield or fail at a critical lateral strain then the whole or part of the monolayer wrinkles and no further axial stress can be sustained. This corresponds to a compressive strain of 0.6% and a maximum axial stress of ~1.1 GPa, which is comparable to the compressive strength of carbon fibres which fail by shear or bulging[20].

5. Conclusions

Micro- and nano-patterning of graphene in specific shapes is fundamental in real device applications. In particular, graphene micro-ribbons have attracted great attention as flexible electronic and optoelectronic devices, and could be potentially exploited as efficient reinforcement in composites. We have produced micro-ribbons of specific dimensions from CVD graphene by combining UV lithography and oxygen plasma treatment. Detailed Raman mapping has proved the effectiveness of the proposed method and has evidenced that the embedded micro-ribbons show a large distribution in the peak positions and FWHM of the G and 2D bands most probably due to structural faults such as wrinkles, folds, substrate surface defects and molecular trapping induced by the fabrication and transferring process. Finally, the mechanical response of the PMMA-embedded micro-ribbons has been investigated by applying uniaxial tensile and compressive loadings to the sample, combined with the simultaneous collection of Raman spectra. The strain rates of both 2D and G peaks have been found in agreement with those typically observed for large membranes of CVD graphene thus suggesting that the proposed micro-patterning procedure does not affect negatively the final mechanical response of the ribbons. Furthermore, the effective Young's modulus of the embedded graphene micro-ribbons has been estimated and found to be 270 GPa, which is in agreement with typical values for wrinkled CVD graphene. The graphene micro-ribbons produced herein are stiffer than carbon nanofibers and comparable in stiffness and compressive strength with carbon fibres. To the best of our knowledge, this is the first time that important parameters such as Young's modulus and compressive strength are reported for the embedded graphene micro-ribbons.

Supplementary Materials: The following are available online at www.mdpi.com/xxx/s1, Figure S1: Characteristic Raman spectrum of the embedded graphene micro-ribbon; Figure S2: Pos(2D) vs Pos(G) for two micro-ribbons.

Author Contributions: Conceptualization, C.G., C.T. and G.T.; Methodology, G.T., E.M. and M.G.P.C.; Data Analysis, A.C.M, G.T. and M.G.P.C..; Writing-Original Draft Preparation, A.C.M, G.T., M.G.P.C. and C.G.; Writing-Review & Editing, A.C.M, G.T., M.G.P.C. and C.G.; Visualization, X.X.; Supervision, C.T. and C.G.; Project Administration, C.G.; Funding Acquisition, C.G..

Funding: The authors acknowledge the financial support of the European Research Council (ERC Advanced Grant 2013) via project no. 321124, "Tailor Graphene". The research project "Graphene Core 2, GA: 696656 – Graphene-based disruptive technologies", which is implemented under the EU-Horizon 2020 Research & Innovation Actions (RIA) and is financially supported by EC financed parts of the Graphene Flagship and the Open FET project "Development of continuous two-dimensional defect-free materials by liquid-metal catalytic routes" no. 736299-LMCat which is implemented under the EU-Horizon 2020 Research Executive Agency (REA) and is financially supported by EC.

Acknowledgments: Dr. John Parthenios and Prof. Kostas Papagelis are acknowledged for useful discussions.

Conflicts of Interest: The authors declare no conflict of interest.